\documentclass[fleqn,usenatbib]{mnras}

\usepackage{newtxtext,newtxmath}

\usepackage[T1]{fontenc}
\usepackage{ae,aecompl}

\usepackage{graphicx}	
\usepackage{amsmath}	
\usepackage{amssymb}	

\usepackage{relsize}  

\title[A new analytic ram pressure profile]
	{A new analytic ram pressure profile for satellite galaxies}

\author[C. A. Vega-Mart\'inez et al.]{
Cristian A. Vega-Mart\'inez,$^{1,2}$\thanks{E-mail: cristian.vegam@userena.cl}
Facundo A. G\'omez,$^{1,2}$
Sof\'ia A. Cora,$^{3, 4}$
Tom\'as Hough$^{3, 4}$
\\
$^{1}$ Instituto de Investigaci\'on Multidisciplinar en Ciencia y Tecnolog\'ia, Universidad de La Serena, Ra\'ul Bitr\'an 1305, La Serena, Chile\\
$^{2}$ Departamento de Astronom\'ia, Universidad de La Serena, Av. Juan Cisternas 1200 Norte, La Serena, Chile\\
$^{3}$Instituto de Astrof\'isica de La Plata (CCT La Plata, CONICET, UNLP), Paseo del Bosque s/n, La Plata, Argentina\\
$^{4}$Facultad de Ciencias Astron\'omicas y Geof\'isicas, Universidad Nacional de La Plata, Paseo del Bosque s/n, La Plata, Argentina\\
}

\date{Accepted XXX. Received YYY; in original form ZZZ}

\pubyear{2021}

\begin{document}
\label{firstpage}
\pagerange{\pageref{firstpage}--\pageref{lastpage}}
\maketitle

\begin{abstract}
We present a new analytic fitting profile to model the 
ram pressure exerted over satellite galaxies 
on different environments and epochs. 
The profile is built using the information of the 
gas particle distribution in hydrodynamical simulations 
of groups and clusters of galaxies to measure the ram pressure directly.
We show that predictions obtained by a previously 
introduced $\beta$--profile model can not consistently reproduce 
the dependence of the ram pressure on halocentric distance 
and redshift for a given halo mass.
It features a systematic underestimation of the predicted ram pressure at high
redshifts ($z > 1.5$), which increases towards the central regions of the 
haloes and it is independent of halo mass,
reaching differences larger than two decades for satellites at 
$r<0.4R_\mathrm{vir}$.
This behaviour reverses as redshift decreases, 
featuring an increasing over--estimation with halocentric 
distance at $z=0$. 
As an alternative, we introduce a new universal analytic model
for the profiles which can recover the ram pressure dependence on halo mass, 
halocentric distance and redshift. 
We analyse the impact of our new profile on galaxy properties 
by applying a semi-analytic model of galaxy formation and evolution 
on top of the simulations. 
We show that galaxies experiencing large amounts of 
cumulative ram pressure stripping typically have 
low stellar masses ($M_\star \leq 10^{9.5} \text{M}_\odot$). 
Besides, their specific star formation histories 
depend on the ram pressure modelling applied, 
particularly at high redshifts ($z > 1.5$).
\end{abstract}




\section{Introduction}

During the last decades, observations of the galaxy 
population inhabiting different environments have shown 
a clear bimodal distribution in several galaxy properties
including colours, morphology, stellar ages and
star formation rates, among others
\citep[e.g.][]{Kauffmann2003, Baldry2004, Cassata2008, Thomas2010, Peng2010, Wetzel2012, Taylor2015}.
This suggests a division between galaxies dominated by 
recently formed stellar populations and galaxies 
with old stellar content.
A critical difference between these populations is
the depletion of the global star formation activity. 
This state, also referred to as galaxy quenching, is generally 
defined when the specific star formation rate (sSFR) of a galaxy, 
i.e. the rate of stars formed  divided by 
its stellar mass, decreases below a certain value, 
being $10^{-11} \text{yr}^{-1}$ the usual threshold 
for galaxies at $z=0$
\citep{Weinmann2010, DeLucia2012, Wetzel2013}.

The contribution to the star formation quenching
from the different physical processes driving galaxy evolution
is  still a topic of debate
\citep[see][for a review on physical models on galaxy formation]{Somerville2015}.
Two different types of processes have been invoked
in the suppression of the star formation: 
mass- and environmental- quenching \citep{Peng2010}.
Comparisons between the properties of populations of
star-forming (active) and quiescent (passive) galaxies 
have shown that, up to  $z \sim 1$, 
it is possible to identify the main 
mechanism driving them to the quenching state
\citep[e.g.][]{Baldry2006, Peng2010, Muzzin2012, Kovac2014, Guglielmo2015, vanderBurg2018, vanderBurg2020}.
This, however, does not necessarily mean
that they are physically unrelated.
At higher redshifts, the picture is more intriguing. 
The median star formation rate (SFR) of galaxies
and the quenched fraction (i.e. the ratio between 
the number of quenched galaxies and the total number of galaxies, 
including quenched and star-forming ones)
are observed to be independent of the environment
\citep[e.g.][]{Darvish2016}.
However, it has also been argued that the environment can 
have an impact on galaxy quenching even up to 
$z \sim 1.6 $ in systems associated to the 
largest overdensities \citep{Nantais2016, Nantais2017}.

Environmental-quenching 
includes all physical mechanisms affecting galaxies according to
the environment where they evolve
\citep[see][for a comprehensive description of these processes]{Boselli2006}.
At the same stellar mass, the measured fraction of quenched galaxies is 
larger for satellite galaxies than for
centrals \citep[e.g][]{Peng2012}, 
and the differences increase strongly with main host halo mass
\citep[e.g][]{Wetzel2012}, and towards the halo
centre \citep[e.g.][]{Haines2015}.
There are indications that the influence of the
environment is stronger for low-mass satellite galaxies 
($M_\star \lesssim 10^{10}\text{M}_\odot$) 
rather than for higher mass ones
\citep{Haines2006, Bamford2009, Peng2010, Roberts2019},
and this has been supported by theoretical analysis 
\citep[e.g.][]{DeLucia2012, Cora2018, Xie2020}.
Besides, environmental quenching is more evident when 
analysing the largest over-densities, where galaxy clusters 
are located ($M_\text{halo} \approx 10^{15}\:\text{M}_\odot$).
Analysis of dense regions in the local Universe have 
shown higher fractions of quenched galaxies 
\citep[e.g.][]{vandenBosch2008, Gavazzi2010}, being the 
sSFR of the satellite galaxies significantly 
smaller in clusters than in lower density regions 
\citep[e.g.][]{Haines2013}, and this behaviour is also present
in high redshift clusters 
\citep[e.g.][]{vanderBurg2018, vanderBurg2020}.

The suppression of the gas accretion onto galaxy discs
\citep[usually referred to as \textit{starvation},][]{Larson1980}
and the loss of gas from the galaxy discs due to the interaction 
with the medium through \textit{ram pressure stripping}  
\citep[RPS,][]{Gunn1972} 
have been the two processes most commonly associated 
to the quenching of satellite galaxies in galaxy clusters
\citep{Quilis2000, Wetzel2013, Muzzin2014, Peng2015, Jaffe2015}, 
and also in the Local Group 
\citep[e.g.][]{Fillingham2015, Wetzel2015}.
A critical difference between the effect of each process is the
quenching timescale inferred from the galaxy properties. 
For starvation, once the feeding of gas to the satellite halts, 
the star formation of the galaxy can be suppressed in
$\gtrsim 4\;\text{Gyr}$ \citep[e.g.][]{Peng2015}, 
whereas RPS of the gas disc could quench the galaxy
in much shorter timescales $\lesssim 1\;\text{Gyr}$
\citep[e.g.][]{Quilis2000, Roediger2005, Steinhauser2016}.
Furthermore, observations of galaxy groups have shown that
the quenching of dwarf galaxies 
($M_\star \lesssim 10^8\:\text{M}_\odot$) is mainly driven 
by ram pressure, whereas starvation dominates quenching
of galaxies with $M_\star \gtrsim 10^8\:\text{M}_\odot$
\citep{Fillingham2015, Wetzel2015}. 
Nonetheless, recent observations indicate  
that environment may have a negligible role in 
the quenching of ultra faint galaxies observed in the 
Local Group \citep{RodriguezWimberly2019}.
The differentiation between these processes becomes 
more complicated when we consider that ram pressure can 
also strip part of the hot halo gas of the galaxy,
without necessarily acting over the disc, 
leading to a starvation scenario
\citep{Bekki2002, Steinhauser2016}. 
Furthermore, another environmental process that can contribute
to the star formation quenching of satellite galaxies is 
the stripping of gas and stars via tidal interactions 
\citep{Merritt1983}. Tidal stripping is, however, considered
as a secondary effect in comparison with RPS
\citep[e.g.][]{McCarthy2008, Font2008, Bahe2015}, agreeing 
with inferences from observational results 
\citep[e.g.][]{Boselli2016}.

Due to the hierarchical growing of the dark 
matter haloes \citep{White1978, Davis1985}, 
a large fraction of the satellites of $z=0$
massive clusters is predicted to have been accreted 
as part of less-massive groups \citep{McGee2009}, thus
galaxy quenching can start in these groups in a 
\textit{pre-processing} stage before falling into larger systems 
\citep[e.g.][]{DeLucia2012, Jaffe2015, Pallero2019, Pallero2020},
Moreover, it has been shown that satellites and central 
galaxies can be pre-processed at several virial radius 
from a cluster centre
\citep[e.g.][]{Bahe2012, Zinger2018, Ayromlou2021}.
Recent analysis have shown a strong dependence of 
the quenching fraction on the intra-cluster medium (ICM) 
density, featuring a specific threshold that determines a 
sharp increase in the quenching efficiency 
\citep{Roberts2019, Pallero2020},
raising the connection between the different environments 
and the ram pressure exerted over the satellites residing in them,
and favouring the importance of RPS 
as a crucial mechanism to understand the general 
galaxy quenching process.

The process of gas stripping from galaxies moving through 
dense environments like galaxy clusters has been explored 
using numerical simulations since the early studies of 
\citet{Lea1976, Gisler1976, Toyama1980,Takeda1984}.
These studies calculated axially symmetric individual stripping 
scenarios with two-dimensional hydrodynamical codes.
The first detailed three-dimensional, time-dependent hydrodynamical 
calculation of an elliptical galaxy orbiting in a gas-rich cluster 
was performed by \citet{Toniazzo2001}. They showed 
that stripping is less efficient than previously reported.
In addition, the first three-dimensional hydrodynamical simulations 
of disc galaxies experiencing RPS on a cluster-like environment 
was presented by \citet{Roediger2007}, showing that the 
mass loss history can not be accurately described by 
the \citet{Gunn1972} analytical estimate.
Different techniques have been applied to study the 
stripping process acting on satellite galaxies within 
groups and clusters with hydrodynamical simulations.
Among these techniques, we found
simulations of isolated galaxies exposed to wind tunnel-like 
mediums to mimic their moving across dense environments
\citep[e.g.][]{Roediger2005, Roediger2006, RoedigerBruggen2006, Kronberger2008, Bekki2009, Tonnesen2010, Tonnesen2012, Ruszkowski2014},
galaxies following orbits throughout galaxy clusters
\citep[e.g.][]{Vollmer2001, Roediger2007, Roediger2008, Jachym2007, Tonnesen2008, HeB2012, Vijayaraghavan2013}, 
and more realistic set-ups of populations of satellites being 
evolved in clusters
\citep[e.g.][]{Vijayaraghavan2015, Jung2018}.
These theoretical analysis have allowed to 
characterise the stripping process of galaxy hot gas haloes
\citep[e.g][]{McCarthy2008, Vijayaraghavan2015},
the mass loss from the galaxy discs
\citep[e.g.][]{Roediger2007, RoedigerBruggen2006, Bekki2009, Bekki2014, Ruggiero2017},
and even the formation and morphological evolution 
of the wakes and tails of stripped mass from discs
\citep[e.g.][]{Roediger2008, Tonnesen2010, Roediger2014, Ruszkowski2014}.
It has been shown that
the external pressure of the ICM can produce 
an enhancement of the star formation in galaxy discs 
\citep[e.g.][]{Kronberger2008, Roediger2014, Bekki2014}
or in bulges
\citep[e.g.][]{Tonnesen2012}.
Moreover, the effect of the inclination angles of galaxy 
discs with respect to their orbital directions across different
environments has been studied
\citep[e.g.][]{RoedigerBruggen2006, Bekki2014},
and it has also been shown that RPS can have
an important effect on the galaxy properties on galaxy groups 
\citep[e.g.][]{Kawata2008, Bekki2009}.
Nonetheless, discrepancies on the stripping rate depending on 
the numerical technique applied to solve the hydrodynamical 
interactions in the gas modelling have been reported \citep{HeB2012}.
Therefore, different strategies to model this environmental 
effect must be considered.

The semi-analytical models (SAMs) of galaxy formation and 
evolution are another useful technique broadly applied to study the 
processes driving the evolution of the galaxy properties 
in a cosmological context. 
In the last years, the use of these models has allowed 
to make important contributions to our understanding of 
the impact of RPS on galaxy properties.
Recent SAMs with updated treatments for RPS 
on satellite galaxies, including gradual stripping 
of the hot gas and mass loss from discs 
\citep{Stevens2017, Cora2018, Xie2020, Ayromlou2021}, 
have shown that RPS acting on both components
is an important ingredient to consistently predict the quenched 
fraction of satellites in different environments.
In addition, \citet{Tecce2010} introduced a  
hybrid technique to model RPS in SAMs when they are applied on 
non--radiative hydrodynamical simulations of galaxy clusters.
This treatment separates
the modelling of the exerted pressure on a satellite
from the mass loss due to the stripping, where the former 
is measured directly from the gas particle distribution of the simulation
and the latter is modelled analytically inside the SAM. 
In this way, the uncertainties regarding to the modelling of the 
interaction between the gas phases 
(i.e. the stripping of the gas from the different 
galaxy components) can be explored analytically.
With this approach they also showed that the assumption 
of an analytical density profile to describe the ICM give 
rise to an overestimation of the RPS at $z>0.5$.
Subsequently, 
\citet{Tecce2011} introduced a fitted analytic profile to 
model the ram pressure on different environments,
from measurements of the pressure in the same set of 
adiabatic hydrodynamical simulations of galaxy clusters.
On the other hand,  \citet{Ayromlou2019}
introduced the Local Background Environment (LBE) technique 
to approximate ram pressure
from dark matter--only $N$--body simulations, based on
a detailed analysis of the particle distribution of the simulation.
Following these type of approaches,
this work is focused on the ram pressure modelling
using analytical estimations obtained from numerical simulations. 
The next subsection describes the modelling of 
ram pressure we consider in our analysis.

\subsection{Ram pressure estimation}

The ram pressure (RP) is the result of the interaction 
between the galaxies and the medium surrounding them,
across which they are moving. 
In case of satellite galaxies, the RP is exerted by
the ICM and is determined by the density of the 
medium in the satellite position and its orbital evolution.
Specifically, it is defined by
\begin{equation}
 P_\mathrm{ram} \equiv \rho_\mathrm{ICM} v^2,
 \label{eq:Pram}
\end{equation}
where $\rho_\mathrm{ICM}$ is the ICM density in the host halo 
and $v$ is the relative velocity between the satellite and the medium.  
Both quantities depend on the galactocentric distance of the satellite.
Therefore, this strong dependence on satellite galactocentric distance
must be accounted for when estimating the amount of
stripped gas mass from the galaxy.
When the gravitational
restoring force of the galaxy is smaller than the 
RP, then the diffuse inter-stellar medium can be stripped
from the satellite
\citep{Gunn1972, Abadi1999, Quilis2000, Vollmer2001, Jaffe2015}.
According to this definition, the estimation of the effective RP 
exerted over a particular observed galaxy requires to define a model 
of the host halo gas density profile, and to estimate the 
orbital galaxy speed relative to the host
\citep[e.g.][]{Rasmussen2008, Jaffe2015, Jaffe2018, Roberts2019}.
When analysing numerical simulations, on the other hand, the estimation 
depends on the type of calculation considered. 

In this work, we use adiabatic hydrodynamical simulations to 
analytically model the RP exerted on satellite galaxies 
inhabiting different environments.
We use the properties of the gas particles  
considered in this type of simulations 
to directly measure the density of the ICM
at the precise positions where the galaxies  
are located by applying the \citet{Tecce2010} technique,
which allows to obtain smooth ICM density profiles in agreement 
with X-ray observations of galaxy clusters
\citep[as is shown in][figure 1]{Tecce2010}.
Thereby, 
these calculations allow to study the evolution of the RP
across cosmic time, to measure its dependence with host halo mass, 
and to define analytic profiles to model it, as done by 
\citet{Tecce2011} (hereafter T11).
These analytic profiles can be considered to consistently model 
the fraction of mass stripped from satellites by RP 
over their evolution with 
semi-analytic models of galaxy formation \citep[e.g.][]{Cora2018}. 
SAMs are based on a set of analytic prescriptions to account for 
the main physical processes driving galaxy formation 
and thus can not self--consistently follow 
the spatial distribution of gas in the galaxies.
Such models are usually coupled to 
dark matter halo properties and merger trees obtained from
cosmological simulations, or created from a
(extended) Press-Schechter formalism
\citep{Press1974, Bond1991, Lacey1993}.

As we have already mentioned, 
there is increasing evidence indicating that RPS is a key process 
in galaxy evolution. Hence, a detailed, accurate and simple modelling of this 
environmental effect is becoming highly demanded. 
Accordingly, 
in this work we revisit the predictions of the analytic 
RP profile presented by T11, 
measured from cosmological hydrodynamical simulations. 
We show that it features important 
systematic differences between its predictions for the 
RP and the expected values measured in 
the simulations considered in its original calculation. 
Moreover, we go further introducing a new analytic profile
able to overcome those differences in the predictions, 
achieving higher accuracy in the modelling.
This paper is organised as follows.
In Section~\ref{sec:t11profile}, we describe the original T11 profile, 
detailing the simulations used in its estimation
and we analyse the problems found in its predictions. 
In Section~\ref{sec:newprofile}, we introduce the new analytic profile 
based on the same simulations and show the improvements 
regarding the predictions. 
In Section~\ref{sec:galprop}, we evaluate the impact of the usage of these
profiles in the galaxy properties by the application of a SAM.
Finally, in Section~\ref{sec:conclusions}, we summarise our analysis 
highlighting the main results.

\section{T11 ram pressure profile}
\label{sec:t11profile}

In T11, a set of N-body/Smooth Particle Hydrodynamical (SPH)
adiabatic resimulations of clusters of galaxies were  
used to measure the RP 
exerted on galaxies and thus to estimate RP profiles
as a function of the halocentric distances, redshifts and the 
host halo virial masses.
The resulting shapes of the RP profiles within these haloes 
were characterised by a $\beta$--profile \citep{Tecce2010},
and a numerical fit of its parameters were provided
as a main result. In the following, 
we describe the simulations used in this work (also considered by T11)
and present an overview and a subsequent analysis of the T11 profile.

\subsection{The simulations}
\label{sec:simulations}

As in T11, 
we use the set of non--radiative N-body/SPH resimulations 
of galaxy clusters described in \citet{Dolag2005, Dolag2009}, 
which were calculated following the entropy--conserving formulation
of SPH \citep{Springel2002} and considering the usual 
parametrization of artificial viscosity
\citep[tagged as \textit{ovisc} in][]{Tecce2010}.
These correspond to 
resimulations of selected regions from a $\sim 685\:\text{Mpc}$ 
side-length volume, characterised by a 
$\Lambda$CDM cosmology with $\Omega_\mathrm{m}=0.3$, 
$\Omega_\Lambda=0.7$,  a Hubble constant $H_0 = 70 
\mathrm{\:km\:s^{-1}\:Mpc^{-1}}$ and a power spectrum with normalization 
$\sigma_8 = 0.9$. The resimulations consider a universal baryonic density
$\Omega_\mathrm{b}=0.039$, with a dark matter particle mass $1.13\times10^9 
h^{-1}\:\mathrm{M}_\odot$ and gas particle mass 
$1.69\times10^8 h^{-1}\:\mathrm{M}_\odot$ \cite[for more details 
about these simulations we refer the reader to][]{Dolag2009}.
The identification of the dark matter haloes and the
construction of their corresponding merger trees were done 
following \citet{Springel2001}. 
Each system of dark matter haloes is characterised by the identification of
a \textit{main host} halo (i.e. the largest overdensity found
over the background), which is detected through a 
friends-of-friends (FOF) technique \citep{Davis1985}. 
The centre of each halo
was defined according to the position of its most bound dark matter
particle.
Properties like the virial mass and radius of the main host
haloes were obtained from the resulting density profiles 
given by the distribution of
the dark matter particles at each redshift.
Assuming a spherical-overdensity criteria \citep{Gunn1972},
we define the virial mass $M_{200}$ 
of each main host halo as
the mass contained in a sphere of radius $R_{200}$ at which the density
equate a $\Delta$ factor of the critical density of the Universe.  
With a constant value of $\Delta = 200$, the virial mass results
\begin{equation}
   M_{200} (< R_{200}) = 200 \rho_\textrm{c} \frac{4\pi}{3}R^3_{200},
\end{equation}
where $\rho_\textrm{c}$ is the critical density of the Universe.
Then, all the haloes lying within each main host 
(also referred to as subhaloes)
were identified using \textsc{Subfind}
\citep{Springel2001}, which defines subhalo particle members 
based on a binding energy criteria.
Excluding the main subhalo, corresponding to the host 
of the central galaxy of each system, all the other
resolved subhaloes are considered as  
satellites of the main host.
Following these definitions, throughout this work,
we use the term \textit{halo} to refer indistinctly 
to any dark matter overdensity, 
independent if it is detected over the background density 
or inside another halo. We specify if we refer to
a main host or a subhalo when needed.

The resimulations considered here are three regions centred in 
large overdensities corresponding to massive galaxy clusters 
with $M_{200} \sim 10^{15} h^{-1}\:\text{M}_\odot$
\citep[originally labelled g1, g8 and g51 in][]{Dolag2005, Dolag2009},
and five regions corresponding to low-mass galaxy clusters
with $M_{200} \sim 10^{14} h^{-1}\:\text{M}_\odot$
(labelled g676, g914, g1542, g3344 and g6212).
These high-resolution regions also include a set of resolved 
smaller clusters and groups in the surroundings of the central clusters.
From those, we select the main host haloes that are
free of particles with lower mass resolution, 
to create a sample of non-contaminated systems 
to be also considered in our analysis. 
Thereby, the analysed sample includes 759 systems in the mass range
$11 < \log(M_{200} h^{-1} [\mathrm{M}_\odot]) < 14$, and 
8 galaxy clusters with $\log(M_{200} h^{-1} [\mathrm{M}_\odot]) > 14$.

\subsection{Ram pressure profile model}

The analytic RP profile introduced by T11
was obtained from the simulations described above, 
using the positions and velocities of the satellites to trace the
RP experienced by them at different galactocentric distances, epochs, and 
ranges of main host halo masses.
Additionally, they included the SAM of galaxy formation and evolution, 
\textsc{sag} \citep{Cora2006, Lagos2008, Tecce2010}, 
only to extend the sample of RP profile tracing points
by incorporating the orphan satellite galaxies, which are  
followed by the semi-analytic modelling. 
These orphans are galaxies that have been hosted by haloes which
are no longer detected by the halo finder because
they have either merged with the main host halo or their
masses decreased below the resolution limit of the simulation.
Therefore, the positions and velocities of these satellites are
defined according to the properties of the most bound particle of the halo
to which each galaxy last belonged \citep[see][]{Tecce2010}.

As was shown in T11, the efficiency of RP increases 
with the mass of the main host halo and with decreasing redshift.
The best--fitting profile was found to be a
full $\beta$--profile, commonly used for characterising the ICM
around massive galaxies \cite[e.g.][]{Churazov2008}. In this case, 
the profile is defined by
\begin{equation}
   P_\mathrm{ram} (M,z) = P_0(M,z) \left[ 1 + \left( \frac{r}{r_\mathrm{s}(M,z)}
   \right)^2 \right]^{-\frac{3}{2}\beta(M,z)},
   \label{eq_Pram_T11}
\end{equation}
where the central value $P_0$, the characteristic radius $r_\mathrm{s}$ and
the exponent $\beta$ are free parameters which depend on both the host
halo mass and the redshift. In T11, these parameters were fitted for
different samples, covering a set of bins in main host halo masses, and for
each simulation snapshot in the redshift range $0 \leq z \leq 3$. 
To obtain the dependence
of these parameters on the redshift, a linear relation in terms of the
expansion factor of the Universe, $a$, was proposed according to
\begin{align}
   \log \left( \frac{P_0}{10^{-12} h^2\,{\rm dyn}\,{\rm cm}^{-2}} \right) &= A_P +
   B_P (a -0.25), \\
   \frac{r_{\rm s}}{R_{200}} &= A_r + B_r  (a -0.25), \\
   \beta &= A_\beta + B_\beta (a -0.25), 
\end{align}
where all the coefficients $A_i$ and $B_i$ with $i=P$, $r$, or $\beta$ depend 
in principle on the main host virial mass. 
Additionally, a linear model following the expression
$a+b(\log M_{200} -12)$ was proposed to parametrize the behaviour of each of
the $A_i$ and $B_i$ coefficients, 
doubling in this way the total number of free parameters involved 
in the fits. The numerical result of these twelve coefficients 
are given by the equations (6a) to (6f) in T11. 
The resulting error associated to each one of them was variable,
reaching a maximum value of $25$ per cent in a couple of cases. 
It was also found that both $A_\beta$ and $B_\beta$ were not dependent 
on the main host halo masses. Consequently
the $\beta$ parameter was expressed only as a function 
of the redshift. Thereby,
all the RP profiles in the analysed redshift range were expressed in
terms of ten numerical constants which define the shape of the profiles 
according to  time and host halo mass for each satellite galaxy 
inhabiting a main host dark matter halo.  

\subsection{Accuracy of the model}
\label{sec:t11comp}

Now we proceed to evaluate the reliability 
of the T11 RP profile model in
predicting the effective RP exerted on satellite galaxies. 
To achieve a complete consistency between our analysis and their analytic RP 
profiles, the same set of hydrodynamical simulations are considered 
throughout our work.
We extract all the positions of the satellite haloes belonging 
to main host haloes detected in the complete resimulated regions described 
in Section~\ref{sec:simulations}, 
excluding only the main haloes contaminated with boundary particles. 
This considerably extend the number of inspected  main host in comparison 
with T11, particularly in the low-mass regime. Furthermore, 
this comparison is restricted to only use the positions of
the satellite haloes identified by the halo finder, 
excluding all the extra tracers considered in T11 derived from the SAM  
(i.e. orphan satellites). 
Although this restriction reduces the number of points to be used 
to define the profiles, it also limits the analysis only to the 
resimulations and their (sub)haloes. 
We can therefore avoid the usage of a SAM in this calculation.

At each satellite position, we measure the effective RP obtained
from the simulation, $P_\text{ram}^\text{\;sim}$, by following the method
described in \cite{Tecce2010}. Here, the ambient density 
and relative velocity with respect to the satellite  
are measured using the gas particles located in a sphere 
centred at the position of the satellite, 
excluding those associated to the subhalo.
Specifically, we remove those neighbouring gas particles 
lying within $2.5$ times the 
radius of the subhalo (as given by \textsc{Subfind})  
having densities larger than twice the median gas density 
calculated from  selected set of particles. 
This filtering procedure is repeated in an iterative 
process until the median density converges, 
thus recovering a smooth ICM without significantly 
affecting the median density profile.
With these estimations, the equation~(\ref{eq:Pram}) is directly applied. 
Besides, the predicted value of RP according to the T11 model,
$P_\text{ram}^\text{\;fit}$, 
is also estimated at each satellite position.

\begin{figure}
   \centering
   \includegraphics[width=\columnwidth]{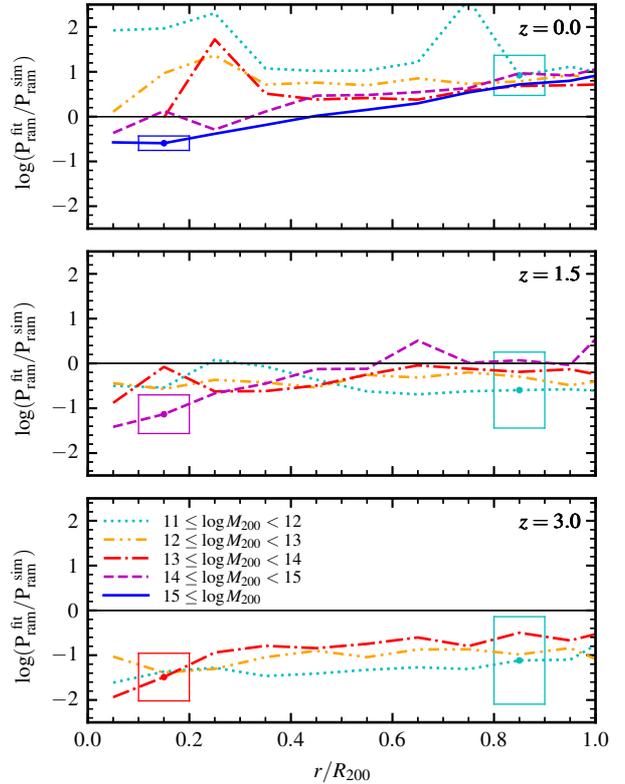}
   \caption{The mean ratio between the RP predicted by the T11 analytic model,
 $P_\text{ram}^\text{\;fit}$, and the RP measured from the gas 
 particles of the simulations, $P_\text{ram}^\text{\;sim}$, in logarithmic scale,
 for $z=0$, $1.5$ and $3.0$ in the \textit{upper}, \textit{middle} and 
 \textit{bottom} panels, respectively.
 The measurements are shown with respect 
 to the galactocentric distance normalised
 to $R_{200}$, grouped according to the instantaneous main 
 host halo mass and shown with different line styles and colours, 
 as detailed in the legend. Empty squares show 
 the size of the chosen bin in the relative distance (horizontal)
 and one standard deviation around the mean (vertical), depicted at
 $r/R_{200} = 0.15$ ($0.85$) for the most (least)
 massive halo sample, respectively.}
   \label{fig:rpcomp}
\end{figure}

The comparison between 
$P_\text{ram}^\text{\;sim}$ and $P_\text{ram}^\text{\;fit}$
is shown in Fig.~\ref{fig:rpcomp} 
for three different redshifts, $z=0$, $1.5$ and $3$, in different panels.
Each one of the samples is separated according to the instantaneous main
host halo mass, and they are indicated with different line styles.
The mean ratios between the predicted and the measured values of the RP
felt by the satellites are shown versus the 
corresponding relative radial distances to the centres of their 
main host haloes, $r/R_{200}$.
As a reference, one standard deviation around the mean 
is also included in the figure through closed rectangles for two selected 
radial distances in each case: 
at $r/R_{200}=0.15$ for the most massive haloes found at each redshift,
and at $r/R_{200}=0.85$ for the least massive ones. This allows us
to compare more precisely the halo masses or redshifts at which 
the difference between the model and the measurements become more significant.
As it can be appreciated from the figure, 
the T11 model overestimates the predicted RP at $z=0$ for more
than 0.5 decades in most of the analysed 
main host halo masses, except for the
most massive ones ($\log(M_{200} [h^{-1} \textrm{M}_\odot]) > 15$) where the
$P_\text{ram}^\text{\;fit} / P_\text{ram}^\text{\;sim}$ ratio 
monotonically decreases towards the centre of the hosts. Here,
the profile underestimates the RP for satellites within 
distances $\leq 0.4 R_{200}$.
Moreover, the ratio exhibits a constant increase with time, leading to
a systematic underestimation of the predicted value of the RP
with respect to the one measured at higher redshifts, 
($z \gtrsim 1.5 $) for all the analysed halo mass ranges.
These differences become more important for the innermost satellites 
($r < 0.5 R_{200}$) of the most massive haloes, where
the underestimation is present at any redshift, reaching 
even two decades at $z \sim 3$.
This drawback of the T11 model is particularly troublesome
if we consider that these satellites experience the highest values of 
RP during their evolution.
Thus, this underestimation can have an important impact on
the amount of gas stripped from the galaxies due to this environmental effect, 
changing the overall amount of gas available to form stars. 

The resulting inaccurate behaviour can be explained 
by analysing the effect of the $P_0$ 
parameter on the prediction. This factor regulates the
zero point of the RP profile
inside the hosts, thus modulating the
maximum value of $P_\textrm{ram}$ reachable towards their centres;
a behaviour strongly dependent on the halo mass. 
Since the fit presented by T11
was constituted by several chi square minimizations, 
starting with the evolution in terms
of the expansion factor of the Universe using linear models, 
the dependence with the mass was inadvertently
softened creating an artificial bias. 
This bias can even be noticed in the upper 
panel of the Fig. 4 in T11, 
in which an example of the fit to the evolution of $P_0$ 
over the expansion factor is shown for a specific main host halo mass range. 
Even though the overlapped
best linear fit follows the measured points 
and is always inside the estimated errors, 
this fit has a clear trend to underestimate the value of 
$P_0$ as redshift decreases, 
reaching differences with respect to the original value greater than one decade.
It is clear that, 
even in that first T11 fitting process, the original trend of the
parameter is lost and the prediction is softened due to the
weight of the measurements in all the considered redshift range. 
Therefore, the biased behaviour of the T11 model was a result of 
the high number of free parameters used 
and the consecutive minimizations performed to
complete the process. The possible degeneracies between the ten parameters
and the high dispersion on the data preclude from finding a predictive model
which can reproduce consistently the highest values of the RP 
inside  massive main host haloes.

\section{New ram pressure profile model}
\label{sec:newprofile}

With the aim of having a more accurate model for the RP 
profile, we 
recreated the original T11 fit, but also tried an alternative analytic 
version by inverting 
the order in which redshift and halo mass dependencies are modelled 
for the $P_0$, $r_\textrm{s}$ and $\beta$ parameters.
However, no improvements in the predictions were 
found with this exercise. Therefore, the profile as proposed by T11 does
not constitute a model consistent for different epochs and halo masses.

In order to find a model with better agreement, 
a different approach must be taken. 
The described limitations forces us to
provide an analytical description of the RP profiles 
capable of reproducing
their behaviour for different halo mass ranges
at a given redshift, using the minimum number of free parameters.
Moreover, due to the need of expressing the temporal evolution of those 
parameters, degeneracies between them must be avoided.

\begin{figure}
   \centering
   \includegraphics[width=\columnwidth]{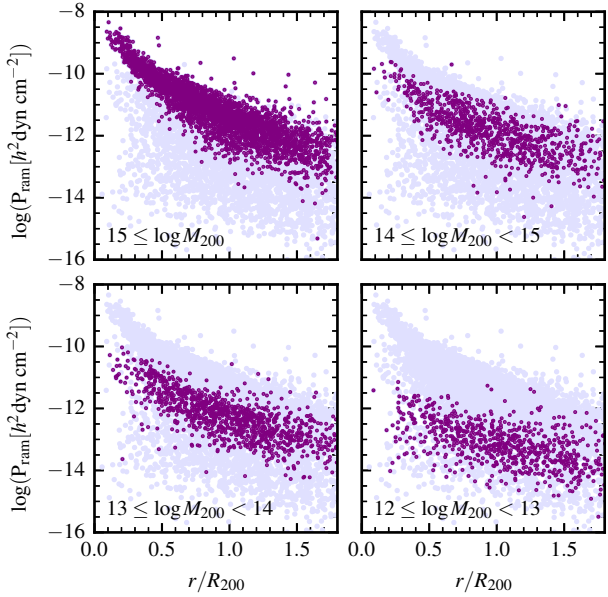}
   \caption{RP values obtained by applying the 
   \citet{Tecce2010} technique on each satellite 
   of the selected haloes from the resimulated regions at $z=0$, 
   as a function of halocentric 
   distance normalized with the virial radius of the halo, 
   for different halo mass ranges.
   For reference, all panels depict the complete set of measurements
   with light purple dots. Each panel highlights, with dark purple dots, 
   the measurements corresponding to the satellites from host haloes 
   selected according to the mass range indicated in each legend.}
   \label{fig:rpbymass}
\end{figure}

An example of the measured RP on the simulations, 
using the \citet{Tecce2010} technique, can be seen in
Fig.~\ref{fig:rpbymass}. Here, the values of $P_\textrm{ram}$ for
the complete sample of satellites found at $z=0$ are shown according to
their relative distances to the centre of the haloes they reside in. In order to
understand the main host halo mass dependence, all measurements are included
in light purple dots, in all the four panels of the figure. 
Each panel highlights, with dark purple dots, the satellites that 
belong to a given mass range, as indicated on the key at the bottom.

According to this, the proposed expression should be able to follow the
steep increment of the RP towards the centre of the massive
haloes, which are the systems imprinting the RP in the 
stronger regime and have a higher impact 
in their star--forming satellites.
The expression should also be able to reproduce simultaneously the 
flatter behaviour observed at different radial distances
for the less massive haloes. 
This can be achieved with a profile defined by
\begin{equation}
        P_\textrm{ram}(M,z) = P_0(z) \left[ 
             \frac{1}{\xi(z)} 
             \left(\frac{r}{R_{200}} \right)
                \right]^{-\frac{3}{2}\alpha(M_{200},z)},
        \label{eq:pram_cvm}
\end{equation}
where $r/R_{200}$ is the relative distance of
the satellite to the centre of the main host halo in terms of the virial radius, 
and $P_0(z)$, $\xi(z)$, and $\alpha(M,z)$ are free parameters to
define the shape of the profile. The $P_0(z)$ parameter, expressed
in units of $h^2\textrm{dyn\;cm}^{-2}$, defines the normalization of the
profile, whereas the dimensionless $\xi(z)$ determines the radial scaling,
both dependent only on the redshift. The power $\alpha(M,z)$ 
encapsulates the dependence on the halo mass $M_{200}$
following a linear relation in logarithmic scale according to
\begin{equation}
        \alpha(M_{200},z) = \alpha_\textrm{M}(z) \log\left(
        M_{200}\;h^{-1} [\textrm{M}_\odot] \right) + \alpha_\textrm{N},
        \label{eq:pram_alpha}
\end{equation}
where $\alpha_\textrm{M}(z)$ and $\alpha_\textrm{N}$ are the free parameters to
set the linear model of the power. To break the evident degeneracy between 
these last two parameters, a fixed value of $\alpha_\textrm{N} = -5.5$
was chosen based on preliminary fits of this model to the data. By doing
this, absolute minimums for the remaining three parameters is guaranteed in
the minimisation processes.

Thereby, by using the equations~\eqref{eq:pram_cvm} and
\eqref{eq:pram_alpha} as a new model for the RP, 
we proceed to fit
the three free parameters in each one of the resimulation snapshots through
chi square minimizations. This is done simultaneously considering all the 
measurements of the RP at each redshift. 
The whole redshift range covered by the main host
haloes of these simulations ($0 \leq z < 7.5$) is considered
to perform the minimizations, 
instead of the restricted range ($0 \leq z \leq 3$) adopted by T11.
Furthermore, each minimization is
performed including all the available range in halo mass
covered by the complete set of resimulations. As described in the
previous section, only the positions of the satellite haloes 
identified by the halo finder
are used in this procedure to define the points to measure
and fit the RP profile.
It is worth noting that we do not apply any restriction in radial 
halocentric distance to the satellites of the systems considered in 
this fitting process. Approximately half of the satellites  
of each main host halo mass range shown in Figure~\ref{fig:rpbymass} 
are lying in the external regions of their corresponding hosts, 
having distances larger than $R_{200}$. The measurements of the 
RP experienced by these more distant satellites 
allow tracing the shape of the profile in the outskirts of the haloes, 
where infalling systems can be affected by the medium. 
Note that our profile is defined only considering the halos 
that are assigned to the host according to the FOF results.

\begin{figure}
   \centering
   \includegraphics[width=\columnwidth]{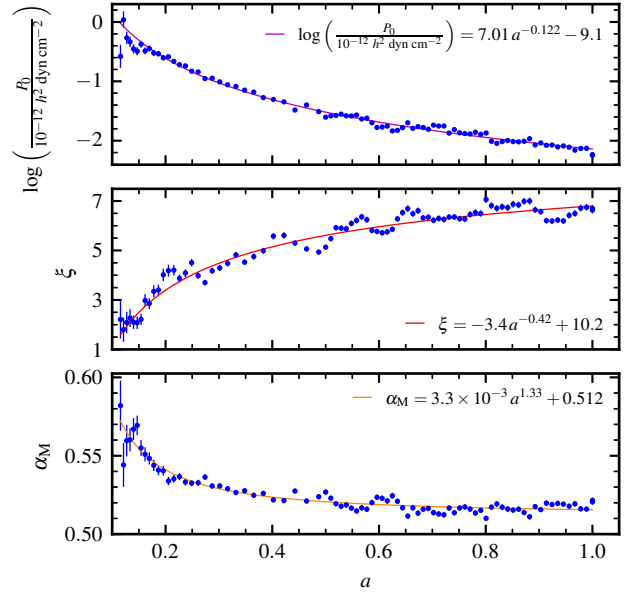}
   \caption{Temporal evolution, in terms of the scale factor $a$, of the 
  $P_0(z)$ (\textit{upper}), $\xi(z)$ (\textit{center}), 
  and $\alpha_M(z)$ (\textit{bottom}) parameters
  of the introduced RP analytic profile. Blue dots show the numerical 
  values found in the fitting process at each snapshot, with the error bars
  extracted from the diagonal elements of the covariant matrices. 
  Solid coloured lines show the fitted model of each parameter.
  The corresponding analytic expression is indicated in each legend.}
   \label{fig:params}
\end{figure}

The resulting low degeneracy between the three
fitted parameters can be seen in the smooth
behaviour of the curves shown in Fig.~\ref{fig:params}. Here, 
the resulting numerical values obtained from all the fits 
of the parameters that characterise the RP profile
at each redshift are shown in blue dots, 
and each snapshot is shown in terms of the 
expansion factor of the Universe $a$. 
The reported errors, included as vertical lines around each dot, 
are extracted directly from the diagonal
elements of the covariant matrices obtained from the fits, assuming no
correlation between them. In the three cases, a very smooth temporal 
evolution of the numerical values is found.
This is exactly the type of behaviour
which allows to recover the desired predictability of the $P_\textrm{ram}$ 
model in the extreme cases of the massive haloes.

Finally, we model the temporal evolution of these parameters through
simple power laws with respect to the expansion factor, 
including a shift in the vertical axis, according to
\begin{equation}
   \mathcal{P} = \lambda_1 a^{\lambda_2} + \lambda_3,
   \label{eq:ccoef}
\end{equation}
where $\mathcal{P}$ corresponds to any of the $\log(P_0)$, $\xi$ and
$\alpha_\textrm{M}$ parameters, 
and the $\lambda_i$ with $i={1,2,3}$ are the three 
free coefficients of the power law model associated to any one of them.
The result of these three fits are also shown in Fig.~\ref{fig:params} 
with solid coloured lines, and clearly smoothly follow  the temporal
evolution of the RP profile parameters (blue points).
The final numerical values obtained for each one of the $\lambda_i$ coefficients 
are listed in Table~\ref{tab:params}, 
together with their respective errors resulting from the chi square
minimization processes. 

\begin{table}
   \centering
   \setlength\tabcolsep{3.0pt} 
   \begin{tabular}{lllllll}
   \hline
   Param. & $\lambda_1$ &$\delta\lambda_1$ & $\lambda_2$ &$\delta\lambda_2$ &
   $\lambda_3$ & $\delta\lambda_3$ \\
   \hline
   $\log(P_0)$ & 7.01 & 3.0 & -0.122 & 0.047 & -9.1  & 3.02 \\
$\xi$ & -3.4 & 1.4 & -0.42 & 0.12 & 10.2 & 1.4 \\
$\alpha_{\rm M}$ & $3.3\times 10^{-3}$ & $1.1\times 10^{-3}$ & 1.33 & 0.18 & 0.512 &
$1.6\times 10^{-3}$ \\
   \hline
   \end{tabular}
   \caption{Resulting numerical values of the $\lambda$ coefficients 
   which describe the temporal evolution of the $P_0$, $\xi$ y $\alpha_{\rm M}$
   parameters defining the RP profile. 
   The uncertainties were extracted directly 
   from the obtained covariance matrices.}
   \label{tab:params}
\end{table}

According to the resulting values of the non diagonal elements of the 
covariance matrices obtained from the fits 
(not included in the table for simplicity), 
we find a degeneracy between the $\lambda_1$ and $\lambda_3$ 
coefficients. This is consistent with the reported errors associated 
to each coefficient, which are smaller than the observed amplitude 
of the little ripples exhibited by the curves. 
Nevertheless, since these values are explicitly
included in the definition of the model, and they are not being fitted
inside another relation, any degeneracy found at this level does not
modify the resulting predictions for the RP model, thus becoming
irrelevant. 

\begin{figure}
   \centering
   \includegraphics[width=\columnwidth]{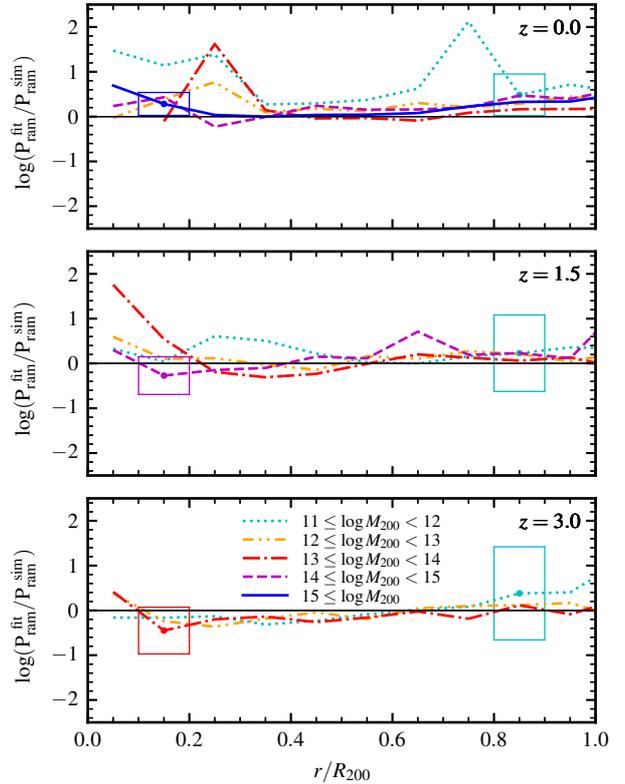}
   \caption{ 
   Comparison between the RP obtained with the new analytic model
   introduced in this work  
   and the one measured from the gas particles of the simulations,
 for $z=0$, $1.5$ and $3.0$ in the \textit{upper}, \textit{middle} and 
 \textit{bottom} panels, respectively. The symbols, binning and 
 selection details are the same as in Fig.~\ref{fig:rpcomp}.}
   \label{fig:rpcomp_new}
\end{figure}

As a proof of the better behaviour of this new model, we
replicate the analysis described in the Section~\ref{sec:t11comp} but 
comparing this new fitted model with the measured RP profile
from the resimulations. This comparison is shown in 
Fig.~\ref{fig:rpcomp_new}. As in Fig.~\ref{fig:rpcomp}, here we show
the dependence of the ratio between the predicted values of RP
given by the model and the measurements from the simulations,
on halocentric distance normalized to $R_{200}$.
It considers the location of satellites from 
different main host halo mass ranges at the same three
redshifts previously analysed. 
According to the
figure, the original reported bias was successfully avoided in this new
model and its accuracy was substantially enhanced.
Most of the artificial dependencies with the relative distance to the centre
and halo mass featured by the T11 are not present in our new profile.
A slight underestimation of the RP towards the 
centre of the most massive main 
host haloes at high redshifts, 
and a general overestimation  
towards the outskirts can be appreciated. 
However, these trends are statistically meaningless considering 
the natural spread of the measured data in the sample of haloes. 

It is appropriate to take into account the dependence of our fitted 
profiles on the numerical method applied to model both the gas dynamics 
and the dark matter evolution in the cosmological simulations. 
Although it has been shown that the technique used to measure RP,
when applied to this particular set of simulations,
is able to recover ICM profiles that are consistent 
with X-ray observation of galaxy clusters \citep{Tecce2010},
it has also been reported that other techniques to model 
the hydrodynamics of the gas can yield differences in the bulk 
gas properties of clusters \citep{HeB2012}. 
Besides, different implementations for the modelling 
of the baryonic physics can affect the 
global properties of the gas content of simulated galaxies
\citep[e.g.][]{Dolag2009, Dave2020}, 
and also the dark matter halo properties
\citep[e.g.][]{BeltzMohrmann2021, Chua2021}.
On the other hand, 
comparisons between different techniques to simulate 
the hydrodynamics in combination with the relevant baryonic processes 
(like star formation and feedback) on galaxy clusters
have shown that the influence of baryons on the
total density profiles can be constrained within the innermost 
regions at $r/r_{500} < 0.01 - 0.1 $, depending 
on redshift \citep{Mostoghiu2019}, 
and they have also shown that
the gas density at $r/r_{500} > 0.2 $ 
depends only weekly on baryon models \citep{Li2020}.
Based on these results, a revision of the fitted free parameters 
included in our proposed profile might be necessary 
to reproduce RP predictions from newer high--resolution
full--physics simulations of galaxy formation, whose predictions
of the hot gas density profiles or 
orbital evolution of satellites substantially differ, 
in particular towards the cluster centres.
Accordingly, in a companion paper \citep{Pallero2020}, we evaluate our  
profile to model RP in galaxy clusters from the
\textsc{C-EAGLE} simulations suit \citep{Barnes2017},
finding a general statistical agreement between the 
spherically averaged RP profiles 
measured from the particle distribution and our analytic model,
without modifying its parameter values.

Accordingly, this new fit is predictive enough to model the RP
at higher redshifts, in contrast to the capabilities of the T11 fit,
and it can be used in different scopes to track the 
amount of gas being lost by satellite galaxies orbiting within their 
main host haloes. 
In the next section, we evaluate the impact of a 
consistent treatment of this environmental effect on satellite galaxies
by comparing the results obtained from applying the new model 
of RP profile and the T11 fit.

\section{Ram pressure in galaxy evolution}
\label{sec:galprop}

In order to analyse the specific effect of RP
on galaxy properties, a galaxy formation and evolution
model must be applied in these simulations, including 
the stripping by RP.

\subsection{Semi-analytic model of galaxy evolution}

To create the galaxy populations from 
the simulations described in Sec.~\ref{sec:simulations},
we consider the updated version of the semi-analytic 
model of galaxy formation \textsc{sag} 
\citep{Cora2006, Lagos2008, Tecce2010, Orsi2014, Munoz2015, Gargiulo2015, Cora2018}.
It uses the halo catalogues and merger trees
to follow the evolution of the galaxy properties, 
assigning one galaxy to each detected subhalo
and solving a set of analytic relations between 
the galaxy components across cosmic time.

Among the included physical processes, the model considers 
the radiative cooling of the hot halo gas, star formation 
(quiescent and in starbursts), and a detailed treatment of the
chemical enrichment considering the contribution from 
stellar winds and different types of 
supernovae \citep{Cora2006}. Thereby,
feedback from these supernovae is also considered. It features an
updated treatment of this feedback 
whose calculation includes an explicit
dependence on redshift based on relations measured from 
full-physics hydrodynamical simulations, 
and it considers ejection of gas from the haloes 
to avoid excess of stellar mass at high redshifts
\citep{Cora2018}. The model
also follows the growth of massive black holes in the centre
of galaxies, and their corresponding feedback which
suppresses gas cooling \citep{Lagos2008}. 
Starbursts can be triggered by galaxy mergers and disc
instabilities, contributing to the formation of 
galaxy bulges \citep{Lagos2008, Munoz2015, Gargiulo2015}.
Environmental effects like tidal and RP stripping 
are also included in the model. It incorporates
a detailed treatment for RPS considering the gas mass loss
from both the discs and the hot gas halo, so that satellite 
galaxies are processed according to a gradual starvation 
scheme of their hot gas \citep{Tecce2010, Cora2018}. 
An additional model to analytically follow  the orbital 
evolution of orphan satellite galaxies is also considered
\citep{Cora2018}, as their positions within the 
main host haloes are relevant for a consistent calculation of
environmental effects in those galaxies. 
Additionally, the free parameters included in the modelled 
relations are usually calibrated to a set of observed 
relations of galaxy properties, by using the 
\textit{Particle Swarm Optimisation} technique \citep{Ruiz2015}.

The gradual starvation scheme to remove the hot gas of
satellite galaxies after infall replaces
the instantaneous removal usually applied in SAMs.
This is a key ingredient to analyse the overall effect of RP
acting on satellites residing in different environments.
The gradual stripping of the hot gas is based on the 
\citet{Font2008} model, considering the estimations 
from \citet{McCarthy2008} 
and assuming a spherical distribution
of the gas. At each timestep, the model calculates a 
satellite--centric radius $r_\text{sat}$ beyond which the
gas is stripped using a dynamic time--scale
\citep[calculated as in][]{Zentner2005}, following the condition
\begin{equation}
    P_\text{ram} > \alpha_\text{RP} 
    \frac{G M_\text{sat}(r_\text{sat})\rho_\text{hot}(r_\text{sat})}{r_\text{sat}},
\end{equation}
where $\alpha_\text{RP}$ is a geometrical constant,
$M_\text{sat}$ is the total satellite mass, $\rho_\text{hot}$ 
is the hot gas density of the satellite and
$P_\text{ram}$ is the RP exerted over the satellite, measured 
directly using the gas particles of the the simulation \citep{Tecce2010}
or estimated from an analytic profile \citep{Tecce2011}.
A general value of $\alpha_\text{RP}=5$ is adopted in the model
\citep{Cora2018}, chosen according to the analyses done by 
\citet{McCarthy2008}.
The calculation of $M_\text{sat}$ considers the contribution of 
the hot gas mass integrated until $r_\text{sat}$, 
using a spherical
isothermal density profile, $\rho_\mathrm{hot}(r) \propto r^{-2}$.

When the ratio between the hot gas mass and the baryonic mass 
of a satellite galaxy decreases below $0.1$, RP can strip gas
from the galaxy disc following the model of \citet{Tecce2010}. 
The stripping radius is calculated by using the \citet{Gunn1972} 
condition
\begin{equation}
    P_\text{ram} > 2\pi G \Sigma_\text{stars}(r) \Sigma_\text{cold}(r),
\end{equation}
where $\Sigma_\text{stars}$ and $\Sigma_\text{cold}$ are the surface
densities of the stellar and gas components of the galaxy disc,
respectively. Both are modelled with exponential profiles with 
the same scale-length, initially calculated from the spin and 
radius of the dark matter subhalo hosting the satellite galaxy.

In both cases of stripping (hot and cold gas), the removed 
mass and metals fractions  are transferred
to the hot gas of the galaxy identified as the central of the 
processed satellite (i.e. the intra-cluster/group medium), which is
the central galaxy of the main host halo in most of the cases.
Besides, it is worth noting that the RPS models included 
in \textsc{sag} do not consider a treatment for the 
stripping of gas from the ejected reservoir 
resulting from supernovae feedback. 
We refer the reader to \citet{Cora2018, Cora2019, Collacchioni2018} 
(and references therein) for more detailed descriptions of
all the physical processes implemented in \textsc{sag},
including the environmental effects considered here.

\subsection{Modelling the galaxy populations}

We applied the galaxy formation model \textsc{sag}
to the complete set of resimulations of clusters of galaxies 
described in Sec.~\ref{sec:simulations}
to trace the evolution of the corresponding galaxy population.
The processing of this particular suit of simulations 
with previous version of \textsc{sag} is described in \citet{Cora2008}, 
\citet{Tecce2010} and \citet{Tecce2011}. 
The model considers the dark matter (sub)haloes 
and their corresponding merger trees as initial and boundary 
conditions to follow the evolution of galaxy properties. 
The calculation of the baryonic content of the subhaloes for
tracing galaxy components  must consider the fraction of mass 
contained in the gas particles.
We recall that, in dark matter--only simulations, the total matter
density of the Universe (i.e. dark matter and baryons) 
is represented by the particle distribution of the simulations. 
Hence, in our analysis when computing total masses of subhaloes, 
for being used by the SAM, 
their dark matter component is corrected by the corresponding 
cosmological baryonic mass fraction.
Therefore, we only use 
the gas particle distribution to measure the 
local RP acting over the semi-analytic satellite galaxies
associated to the corresponding dark matter haloes, 
by applying the \citet{Tecce2010} technique.

We create a fiducial galaxy population applying the full-physics
\textsc{sag} model described in \citet{Cora2018}.
It considers the values of the free parameters that were 
calibrated to a dark matter only simulation of 1 Gpc$/h$ of sidelength
\cite[\textsc{sag}$_{\beta 1.3}$ variant detailed in][]{Cora2018, Cora2019}
instead of tuning a new set of parameter values for this new
resolution limit and cosmology.
The calculation of RPS considers the values of RP measured
from the gas particle distribution of the simulation, 
$P_\text{ram}^\text{\;sim}$. 
Besides, to isolate the effect of RPS, we disable 
the mass loss of gas and stars due to tidal stripping processes.
To validate this model variant, we inspect 
several galaxy properties 
in order to guarantee the consistency of the obtained 
galaxy population, including the total baryonic fraction 
of all the galaxies, the conditional stellar mass function 
and the conditional luminosity function (CLF) of satellites.
The resulting CLFs feature a 
decreasing behaviour with an excess towards
the fainter-end of the function,
when using the \citet{Cora2018} parameter values 
in comparison with the results from \citet{Lan2016}. Therefore, 
we increase the parameters related with the efficiencies of
supernovae feedback to improve the CLFs shape.
Although the break and slopes of the CLFs are recovered 
with this change, the simultaneous fitting of the normalization 
for all the main host mass ranges requires a complete exploration of the
set of model parameters. Hence, as we aim to 
compare galaxies in specific main host mass ranges,
we notice this caveat and proceed with a 
fair model-to-model comparison restricted to each main host mass range.
Besides, a complete recalibration of the current model,
specifically focused on this set of resimulations, 
constitutes a computationally 
demanding task that is outside the scope of this work. 

Subsequently, we create two additional variants of the galaxy modelling by 
considering different analytic estimations of RP:
the fit of the RP profiles introduced by \citet{Tecce2011};
and the revisited fit presented in this work, 
described and analysed in the previous section.

As the three galaxy models are identical except for the RP calculation,
the resulting number of satellite galaxies obtained for each run are 
equal due to 
the \textsc{sag} processing strategy.
The model applies a pre-processing to the halo merger trees in 
order to identify the merging haloes producing orphan galaxies, 
it calculates their orbits and defines their merging timescales.
\citep[for a more detailed description of this pre-processing, 
we refer the reader to][]{Delfino2021}. 
This pre-processing is completed 
before the properties of the stellar and gas content of the galaxies
are calculated.
Therefore, a direct comparison between the satellite populations 
of the same clusters can be done. 
Besides, it is worth noting that when comparing conditional 
luminosity and mass functions of satellite galaxies 
within main hosts in different
mass ranges, the resulting number counts show no 
significant differences between the three model variants 
analysed throughout this section.

\subsection{Mass stripped by ram pressure}

To understand the general impact of the RP modelling, 
we create two samples of satellite galaxies according to 
the mass of the main host halo they belong to.
A massive sample, 
containing  satellites of  simulated clusters with
$\log (M_{200} h^{-1} [\mathrm{M}_\odot]) > 15$. 
This sample contains the three main host haloes 
of the simulations g1, g8 and g51 (see Section~\ref{sec:simulations}).
Satellites of this sample are expected to experience the 
strongest RP values, especially near their cluster centres. The second sample
is composed of all the satellites that belong to main hosts haloes with
$ 13 \leq \log (M_{200} h^{-1} [\mathrm{M}_\odot]) < 14$, selected from 
the full set of resimulations. Considering only the non-contaminated haloes,
this less massive sample includes the satellites belonging to
34 main host haloes.

\begin{figure}
   \centering
   \includegraphics{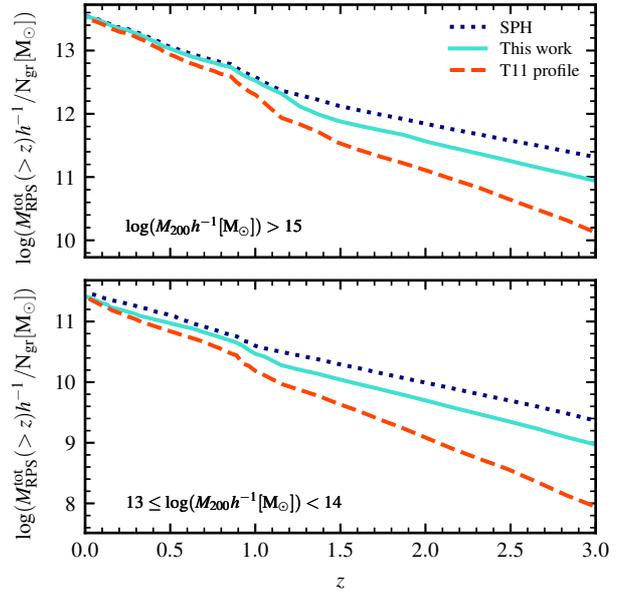}
   \caption{
   Mean cumulative function of stripped mass of the sample of
   satellite galaxies until a redshift $z$.
   The three model variants are depicted with different line styles 
   and colours: the fiducial model in dotted dark blue lines, 
   the model using the new RP profile in solid cyan lines, and
   the model using the T11 profile in dashed orange lines.
   Each function considers the complete population of satellites
   belonging to different ranges of main host halo: satellites from 
   the 3 massive clusters (\textit{top panel}) and from the 
   selected 34 less massive main hosts with 
   $13 \leq \log(M_{200} h^{-1}[\mathrm{M}_\odot]) < 14$ 
   (\textit{bottom panel}).}
   \label{fig:gasstrip}
\end{figure}

We proceed to analyse these two samples in 
each of the three model variants. 
For an overall measuring of the RP effect,
we take into account all the progenitors of the galaxies belonging
to each $z=0$ selected halo 
(i.e. the complete set of members of their merger trees)
and measure the total amount of gas mass
stripped by RP at each simulation timestep.
This allows to define the cumulative function of stripped mass,
$M_\mathrm{RPS}(>z)$, 
as the sum of all the stripping measurements 
from the progenitors of a selected galaxy until a given redshift, $z$.
It includes the gas stripped from both the gas disc and  
hot gas halo components 
to quantify the complete effect of this environmental process. 
As a result, $M_\mathrm{RPS}(>z=0)$ corresponds to the total mass 
that a galaxy has lost by RPS until $z=0$, 
accounting all its progenitors. 
Combining the results obtained from all the selected galaxies, 
we calculate the cumulative 
function of stripped mass of the complete sample,
at any redshift, as the sum
\begin{equation}
M_\mathrm{RPS}^\mathrm{tot}(>z) = \Sigma_i M_\mathrm{RPS}^i(>z),
\end{equation}
where $i$ refers to the selected galaxies of the sample.
Finally, we obtain the mean by dividing by the number of main host
haloes selected in each sample, $\text{N}_\text{gr}$.
The resulting function is shown in Fig.~\ref{fig:gasstrip}.
The top panel shows the results 
of the satellite sample from the higher main host mass, 
whereas the bottom panel 
shows the results for the lower mass sample. 
Each panel shows the three model variants: 
the dotted dark blue lines show our fiducial model considering
the RP as measured from the SPH simulation,
the solid cyan lines show the model using our new analytic RP profile,
and the dashed orange lines show the model using the T11 fit. 

For both mass ranges, the total amount of gas mass stripped tends to reach
the same value at $z=0$, independently of the RP model.
The values are approximately equivalent to the 
2.3 and 0.8 per cent
of the mean main host total mass for the high and low mass samples, 
respectively. Therefore, this global
stripping process seems sightly more significant towards 
higher cluster masses; an expected behaviour as the RP 
acting on satellites near the centre of their hosts is able to
reach larger values when the mass of the main host is high
\citep[e.g.][]{Wetzel2012, Haines2015}.
Nonetheless, it should be warned that the effective 
final RP stripped mass amount has
a direct dependence on the general modelling of galaxy evolution,
i.e. the physical processes considered in the semi-analytic model
and the corresponding calibration of its free parameters.
Besides, as the number of satellites and total mass of each system
increases over cosmic time, the amount of mass involved in newer stripping 
events becomes more significant than the older ones in this figure. 
This also contributes to the similar behaviour of the accumulated 
RP stripped mass exhibited by the three model variants.
On the other hand, the small knee near $z \sim 1$ featured 
by all models only appears when using the larger parameters related 
with the supernovae feedback process in the SAM, 
but is not present with the original 
calibrated parameter set. Thus, this break is
a direct consequence of the physical relation introduced 
to model the supernovae feedback, as its efficiency has a strong dependence 
on redshift and changes its dependence on the halo virial 
velocity when this property reach $60\;\text{km/s}$ 
\citep[see][equations (10) and (12)]{Cora2018}.

The general trends of the two mass bins are analogue. The model which
applies the RP profile described in this work
is more effective in stripping
mass than the one described by T11, particularly at high redshift. 
We note that, although our proposed RP profile represent 
a significant improvement with respect to T11, there are still 
some systematic differences with respect to the fiducial model.
The diversity in the dynamics of the infalling satellites plays a 
role here. Although our analytic model is able to reproduce the 
increasing RP values towards the centre of the haloes 
(see Fig.~\ref{fig:rpcomp_new}), 
a small fraction of satellites with larger 
galactocentric distances can also exhibit 
comparable RP values than those near the centre
(as it can be observed from the vertical spread of the measurements 
shown in Fig.~\ref{fig:rpbymass})
due to their high velocities relative to the host
\citep[e.g. see][]{Oman2013}.
Consequently, the RP felt by these objects can not
be reproduced by the analytic models, as these 
satellites are outliers of the mean trends.
Besides, our fitting profile assumes a spherically symmetric 
distribution of the gas within main
host haloes, so that large values of
RP resulting from inhomogeneities are not considered. 
According to this, the analytic 
fit constitutes a good approximation to the median values of RP, 
but it must be noted that a fraction of the extreme cases of 
stripped galaxies can not be fully recovered with this method.
This is in line with some of the expected limitations associated 
with semi-analytic modelling of galaxy evolution.

Besides these extreme cases, the new modelling of RP profiles
can have a non-negligible impact on the 
properties of the global population of satellites, 
in particular on satellites residing in
high density environments where the median RP is higher.
Therefore, the following analysis will be focused in the 
properties of this galaxy population.

\subsection{Ram pressure on cluster galaxies}

\begin{figure}
   \centering
   \includegraphics{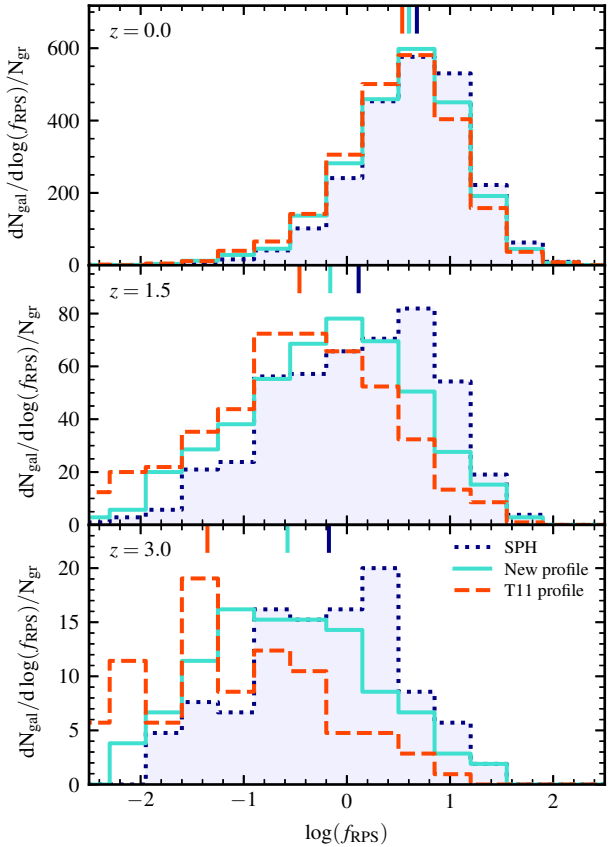}
   \caption{Number counts of satellite galaxies having different values
   of their instantaneous fraction of total stripped mass, defined by
   equation~\eqref{eq:frps}. Each panel shows a different redshift: 
   $z=0$, $1.5$ and $2$ in the \textit{top}, \textit{centre} 
   and \textit{bottom} panels, respectively. Fiducial model resulting 
   distributions are depicted with dotted dark blue lines and
   shaded areas, whereas the new and T11 profiles are depicted 
   with solid cyan and dashed orange lines, respectively. 
   Median values are shown with small vertical lines at the 
   top of each panel using the same colour coding.
   }
   \label{fig:histfprs}
\end{figure}

To quantify the effective impact of the RP on individual galaxies, 
we define the instantaneous fraction of the total stripped mass, 
$f_\mathrm{RPS}$, as the ratio between 
the total stripped mass
of a satellite and its stellar mass, i.e.
\begin{equation}
    f_\mathrm{RPS} \equiv \frac{M_\mathrm{RPS}(>z=z_i)}{M_\star(z_i)},
\label{eq:frps}
\end{equation}
where $M_\text{RPS}(>z=z_i)$ is the 
total stripped mass of a satellite
until a redshift $z_i$, as defined in the previous section, 
and $M_\star(z_i)$ is its instantaneous stellar mass at $z_i$.
The $f_\mathrm{RPS}$  distributions 
exhibited by the complete sample of satellites
belonging to the three most massive clusters are analysed by 
counting the number of galaxies having
different values of this quantity per cluster unit.
To analyse the distributions at different redshifts, 
we consider the  progenitors of the selected
$z=0$ main host haloes, 
and calculate $M_\text{RPS}(>z=z_i)$
for all their satellites using 
their respective merger trees, accordingly.
Hence, $\text{N}_\text{gr} = 3$ in this analysis. 
At each redshift in the fiducial 
model, we select all satellites
having $\log(M_\star[\text{M}_\odot]) \geq 8$, 
being this limit chosen according to the 
behaviour of the conditional stellar mass functions 
of satellites at different redshifts which start 
to misbehave for stellar masses below
$\sim 10^7 \text{M}_\odot$.
We can confidently consider this limit due to 
we are analysing satellite galaxies from a model 
that includes a consistent treatment for the orphan
galaxies, so we can compare low mass satellites 
processed by the environmental effects.
We then identify the haloes of this selected galaxies
and use them to identify
the corresponding galaxies in the remaining two models  
for a direct comparison.
The resulting function is shown in the Fig.~\ref{fig:histfprs}. We 
consider three distinct epochs,
$z=0.0$, $1.5$ and $3.0$ shown in the top, middle and bottom panels, 
respectively. The selection of redshifts considers
the same values and time range as the analysis 
done in the previous sections. 
The distribution of number counts resulting from the 
fiducial model using the values of
RP measured from the SPH simulation is shown with 
the filled area and the dotted dark blue line, whereas the 
solid cyan line and the dashed orange line depict the  
number counts of the model applying our new profile and the 
model using the T11 fit, respectively (as in Fig.~\ref{fig:gasstrip}).
In addition,
the small vertical lines located in the 
upper axis of each panel indicates the median values 
of each distribution.
These medians clearly follow the same trend as 
the cumulative functions of stripped mass
shown in Fig.~\ref{fig:gasstrip},
where lower fractions at any epoch result 
when analytic RP profiles are applied. 
In particular, the T11 fit produces noticeably lower
median fractions as redshift increases, reaching 
differences of almost $1.5$ decades at $z=3.0$ 
with respect to the fiducial model results. 
Moreover, the global shape of the distributions of the 
model applying the T11 profile 
are skewed to lower values of $f_\text{RPS}$ as
redshift increases, indicating a persistent smaller 
fraction of stripped mass in the complete satellite 
galaxy population. The analytic fit presented in this work 
features a noticeable improvement in 
the modelling of the number of galaxies 
with large fractions of stripped mass although, as previously discussed, is
unable to fully recover the total number of galaxies with
larger fractions.

\begin{figure}
   \centering
   \includegraphics{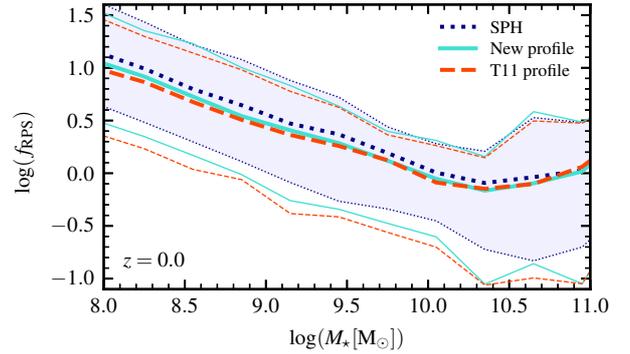}
   \caption{Median values of the instantaneous fraction of total 
   stripped mass of the satellite galaxies of the massive clusters 
   at $z=0$, as a function of their stellar mass. Fiducial model 
   results are depicted with dotted dark blue lines and
   shaded area, whereas the new and T11 profiles are represented by
   solid cyan and dashed orange lines, respectively. 
   Median values are shown with thick lines, whereas the shaded area
   and thin lines depict the percentiles 10 and 90 of the measurements 
   in each stellar mass bin.
   }
   \label{fig:frpsmstar}
\end{figure}

An interesting feature of the distributions 
shown in Fig.~\ref{fig:histfprs} 
is the global increasing mean values of $f_\text{RPS}$ 
over time, exhibited by the three variants of the
galaxy formation model.
As redshift decreases, there is an increasing number of 
satellite galaxies whose accumulated total mass lost by RPS 
is larger than their instantaneous stellar mass 
(i.e. $\log (f_\text{RPS}) > 0$).
This is interesting considering that the stellar mass of these 
galaxies is not being reduced by external processes (like 
tidal stripping) besides  stellar evolution. 
To further analyse the behaviour 
of $f_\text{RPS}$, in Fig.~\ref{fig:frpsmstar} 
we show the median values obtained from the $z=0$ 
satellite sample, considering the three model variants. 
Satellites are binned according to their stellar masses.
The same colour coding as the previous figure is used, but
here the shaded area and the thin lines depict the 
percentiles 10 and 90 of the measurements in each 
stellar mass bin. 
In general, the median $\log(f_\text{RPS})$ shows a constant increase 
with decreasing stellar masses for galaxies having
$\log(M_\star[\mathrm{M}_\odot]) \lesssim 10.4$, 
mass above which the trends feature a turn-up and 
start to increase towards higher stellar masses. 
It is interesting to note that the change in the trends
is parallel in the three model variants. However, 
we can not draw robust conclusions from this higher mass
range as the number of satellites having 
$\log(M_\star[\mathrm{M}_\odot]) > 10.4$
constitute less than a 0.09 fraction of the selected sample 
in each model variants. Furthermore, their measured 
$f_\text{RPS}$ show a significant dispersion. 
In general, we find that galaxies featuring the largest
values of $\log(f_\text{RPS})$ have low stellar masses, 
in the mass regime corresponding to dwarf galaxies.
Therefore, these galaxies are more susceptible 
to be affected by RP, in agreement with 
previous results obtained with the model \citep{Cora2018,Cora2019}, 
and other studies
\cite[e.g. see ][]{Roberts2019}.

To go further in the analysis, now 
the main properties of the population of satellites with 
the largest fractions of stripped mass are compared. 
At each redshift, a cut in $f_\text{RPS}$ 
following the median value obtained from the fiducial model
is chosen to select the galaxies.
The same cut applied to the three model variants allows us to compare 
galaxies that are being affected with analogue stripping processes.
Thereby, differences in the number counts of galaxies allow measuring 
the overall effect of using the analytic models to estimate  RP. 
Median values, indicating the half of the distributions, are
preferred instead of a larger fraction only for visualisation purposes. 
This guarantees a number of galaxies large enough 
to be compared in all considered redshifts.
It is worth noting that trends and general conclusions reported
in this section are not affected when the cut applied in the 
fraction is increased (e.g. using the value defining the 
highest quartile of the distribution instead of the median).
Therefore, the chosen limits in $f_\text{RPS}$
to select the galaxy samples are
$4.78$, $1.30$ and $0.67$ for  redshifts 
$0.0$, $1.5$ and $3.0$, respectively. 

\begin{figure*}
   \centering
   \includegraphics{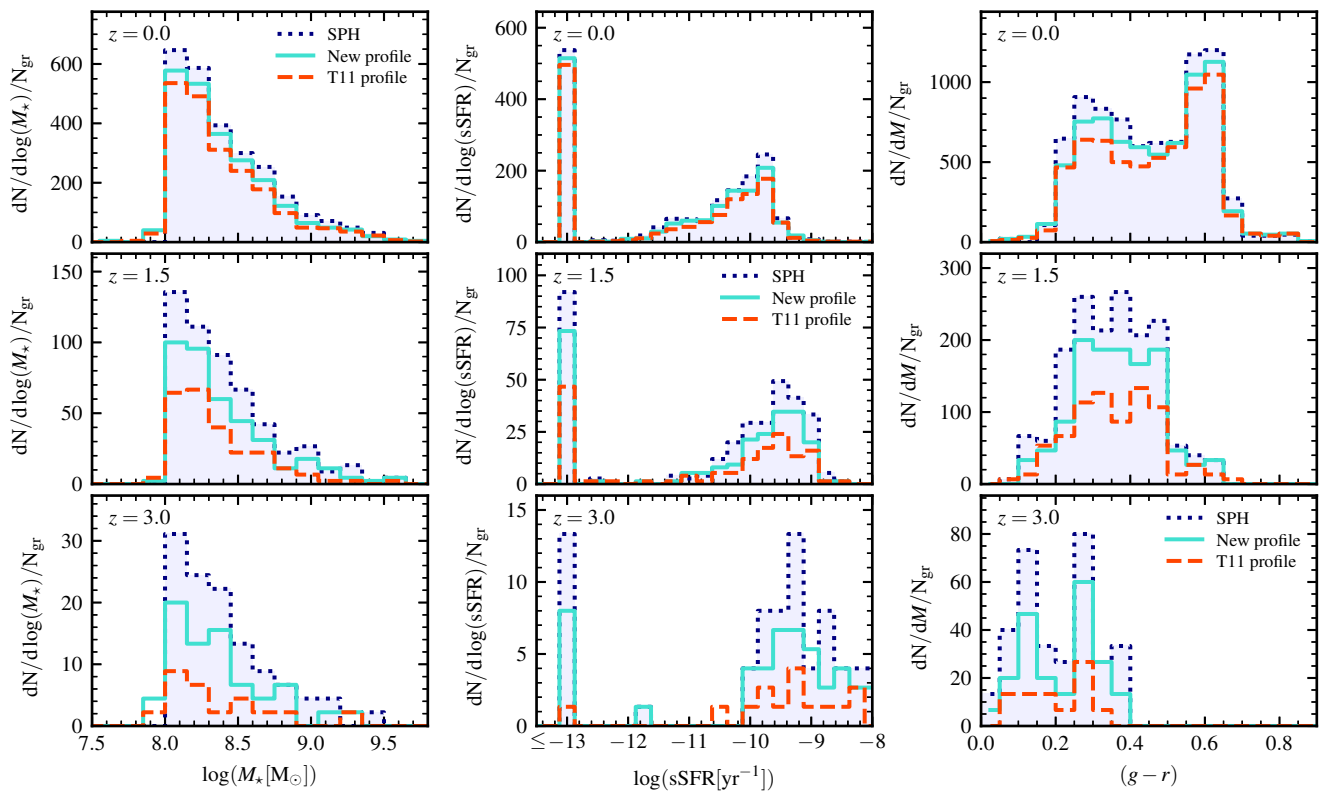}
   \caption{Differential number counts of galaxies binned by: stellar mass 
   (\textit{left} panels), specific star formation rate (\textit{centre} panels),
   and $(g-r)$ colour (\textit{right} panels). Three different epochs
   are considered: $z=0$ (\textit{top} panels), $1.5$ (\textit{middle} panels),
   and $3.0$ (\textit{bottom} panels). At each redshift, the sample of
   galaxies is selected by applying a cut in their
   instantaneous fraction of total stripped mass, $f_\text{RPS}$.  
   The selected value for this cut is chosen according to the median 
   in the distribution of $f_\text{RPS}$
   as found in the model of RPS measured from the SPH simulation:
   $4.78$, $1.30$ and $0.67$ for  redshifts 
   $0.0$, $1.5$ and $3.0$, respectively.}
   \label{fig:galprop50}
\end{figure*}

The comparison between the selected galaxies properties
from the three model variants is shown in Fig.~\ref{fig:galprop50}.
The differential galaxy number counts according to their stellar mass, 
specific star formation rate and $(g-r)$ rest-frame colour  
are shown in the left, middle and right columns, respectively.  
The comparison includes the three considered redshifts 
in different rows, as indicated on each panel. 
The colour coding of the models is the same 
as in Fig.~\ref{fig:histfprs}.
The general trend already spotted before  
remains in all the cases presented here. This is, 
the number of galaxies having the largest amounts
of stripped mass are lower in the models using the analytic profiles
of RP than in the fiducial one, being the T11 profile 
the one featuring the lowest amount of galaxies with large 
stripped mass fractions. 
The distributions of number counts of galaxies
with different stellar masses
show that most of the satellite population 
experiencing the larger fractions of total stripped mass, 
and not recovered when the analytic fits are applied, 
tend to have low stellar mass
($M_\star \leq 10^{9.5} \text{M}_\odot$),
in the regime of dwarf satellite galaxies.
This trend is noticeably larger at high redshifts for the T11 model, 
in agreement with the reported systematic 
strong subestimation of the general
$P_\text{ram}$ modelling exhibited by that profile.
As redshift decreases, the relative impact
of the RPS process into the final stellar mass distribution seems to 
become milder as the satellite mass distributions 
progressively become more similar for the different models. 
Besides, the total number of cluster satellites 
at $z=0$ is considerably larger than at
high redshifts,
and a significant fraction of these satellites 
is expected to be accreted as part of less massive groups 
\citep{McGee2009, Pallero2019, Pallero2020}. 
Hence, the stripping of gas experienced by 
these galaxies has a non negligible contribution from the
environment they inhabited before being accreted by the cluster.
On top of that, the small tail in the low mass limit, 
featured by the models using the analytic RP profiles, 
points towards a general decrease of the resulting galaxy masses, 
in comparison with the fiducial model results.
Moreover, as we are comparing the galaxies with the largest 
amounts of stripped mass obtained with each model variant, 
the trends exhibited at each redshift 
suggest that the general effect of the stripping process 
has a stronger impact on low--mass galaxies, 
in agreement with results derived from observed galaxy properties
\citep{Haines2006, Bamford2009, Peng2010, Roberts2019}.

As the stripping of gas from discs by RP 
is characterised by short time-scales \citep[e.g.][]{Foltz2018}, 
in Fig.~\ref{fig:galprop50} we also compare the  galaxies sSFR 
and $(g-r)$ colour number count distributions obtained from 
the three model variants,  
to analyse the more instantaneous impact of this environmental
effect on their properties.
The \textsc{sag} model does not consider
star formation triggered by gas compression occurring during
stripping events, so galaxies with high SFR in the sample have
their star formation strictly driven by internal processes 
or galaxy mergers.
The sSFR of each galaxy is calculated considering all 
newly formed stars in the time elapsed between two consecutive 
snapshots of the simulation, approximately $200$ Myrs,
divided by their stellar mass. 
All the quenched satellites having values of sSFR
lower than $10^{-13}\;\text{yr}^{-1}$ 
are manually included in the lowest bin of the distributions 
as indicated in the corresponding 
axis label of the figure for visualisation purposes.
The selected sample
shows distributions that continuously tend
to have lower star formation as redshift decreases.
Thereby, using the commonly applied $z=0$ threshold, 
$\mathrm{sSFR} = 10^{-11}\;\text{yr}^{-1}$  
\citep[e.g.][]{Wetzel2012}, as a simplistic approach 
to separate star--forming from quiescent galaxy 
populations on the three analysed redshifts
\citep[as in][]{Fontanot2009}, 
the resulting number of quenched galaxies experiencing 
the largest gas stripping increases over time.
It is however important to note that this threshold  
does not take into account the evolution of the sSFR
distribution over redshift, 
being unable to properly separate the quenched population at $z>0$
\citep[e.g.][]{Fang2018, Donnari2019}. 
As a result, our chosen threshold only serves as a reference value for comparing number 
counts of galaxies in different sSFR regimes. 
According to this, the resulting number of quiescent galaxies 
exhibits the most important differences among the model variants, 
being the model using the T11 profile the one that features 
the largest lack of quenched galaxies at higher redshifts 
in comparison with the fiducial galaxy sample.
Nonetheless, like the distributions of stellar mass, the 
resulting distributions of sSFR at $z=0$ of the three model 
variants are similar. The observed differences at high redshifts,
resulting from the RP modelling, are masked at $z=0$ because of
the increased number of satellites. Besides, 
as a fraction of these satellites are accreted within 
less massive groups, their star formation histories may
have been pre-processed in different environments
outside the virial radius of the cluster \citep{Pallero2019},
leading to different stripping scenarios during their evolution.
On the other hand, as redshift increases,
the difference in the number of both quenched and star-forming 
galaxies becomes more important between the model variants. 
According to the results from the selected sample of galaxies,
the star formation rate of
satellites that have experienced large amounts of RPS 
is highly affected by this environmental effect at
higher redshifts, even until $z \sim 3$, 
as the number of quiescent satellites substantially 
differs between the different approaches of RP modelling.
It is worth noting that 
when the sSFR comparison at $z = 3.0$ is shown
instead using normalised functions, 
the resulting distributions feature the same trend 
spotted here, that is
the fraction of galaxies having 
$\mathrm{sSFR} \leq 10^{-13}\;\mathrm{yr}^{-1}$ 
is significantly lower when the T11 RP profile is applied
in the calculation of the stripping mass.
It has been previously stated that the role of environment 
on galaxy quenching in high density regions 
becomes important only after
$z \sim 1.6 $ \citep{Nantais2016, Nantais2017}.
This, however, does not disagree with the evolution 
of the sSFR exhibited by our sample 
as these galaxies are chosen according to 
their ratio of stripped mass,
which determines a biased selection. 
Besides, as the sample is not compared with analogue field galaxies, 
the environmental quenching efficiency 
\citep{Peng2010, Peng2012}, usually estimated to evaluate the
role of the environment on galaxy quenching, 
can not be calculated using this suit of resimulations.

The distributions of galaxy number counts of 
rest-frame $(g-r)$ colours of the selected galaxies
are shown in the right panel of Fig~\ref{fig:galprop50}. 
As in the case of
the sSFR, differences in galaxy colours allow to evaluate 
the more instantaneous effect of RP in galaxies. 
The $z=0$ colours of the sample exhibit a 
clear bimodal distribution, with a redder peak 
slightly shifted $\sim 0.15\;\text{mag}$ in comparison 
with the usually observed \textit{red} population of galaxies
\citep[e.g][]{Bell2003, Taylor2015}. However, as we are 
using a shallow calibration of the parameters of the
\textsc{sag} model, the global reported colours of the galaxies
could be affected by a systematic shift. 
The distributions nonetheless show a clear evolution across 
redshift, being the galaxies affected by high RP
more star-forming and, consequently, bluer
with increasing redshift.
Moreover, no difference indicating preferred colours are 
found among the different models. The lack in the number 
of galaxies resulting from the analytic models of RP 
is observed in the complete range of colours, being 
this galaxy property insensitive to the RP modelling. 

According to these results, 
our new profile for modelling the RP exerted by galaxy groups and
clusters exhibits a significant improvement in the general RP calculation
in galaxy formation models, 
specifically for high redshift galaxies. 
The number of galaxies experiencing large amounts of RPS 
depends strongly on the RP modelling, and a smaller 
number of these galaxies are obtained with the T11 profile.
Major differences are also found in galaxy properties,  
particularly in their stellar mass and star formation rate. 
We remark that these differences must be taken as a lower 
limit as the \textsc{sag} model does not consider star formation 
induced in the stripped gas by RP, therefore such
new populations of stars formed in 
these galaxies are missing 
in our analysis, a feature commonly observed in 
extremely stripped galaxies \citep[e.g.][]{Poggianti2016}.
Further analysis focused in this specific set of galaxies
using improved simulations of galaxy clusters will allow to 
study more precisely the effect of the stripping 
events on global galaxy properties.

\section{Summary and conclusions}
\label{sec:conclusions}

Using a set of SPH resimulations of galaxy clusters, 
we have analysed different approaches to 
model analytically the RP experienced by satellite 
galaxies inhabiting main host haloes whose masses span 
from less massive groups ($M_{200} \sim 10^{13}\;\text{M}_\odot$)
to galaxy clusters ($M_{200} \sim \text{few}\;10^{15}\;\text{M}_\odot$).
We have also  
considered a large redshift range, from $7.5 > z \geq 0.0$.
The T11 RP profile \citep{Tecce2011} was revisited in detail,
showing misleading predictions at different epochs in 
comparison with the effective RP measured from the
simulations using the \citet{Tecce2010} method. In addition, 
a new universal analytic model for the RP was introduced, 
and the impact of the application of this type of treatment in 
galaxy formation models was evaluated, focusing
specifically on satellites residing in high density environments. 
The main results of these analysis can be summarised as follows:

\begin{itemize}
    \item The T11 analytic RP profile 
    features a systematic underestimation of the RP measured 
    in the simulations where the profile was defined, reaching 
    even 2 decades at $z=3$ in all the analysed main host masses.
    This underestimation persists towards the centres of massive 
    galaxy clusters at $z=0$, however it reverses with increasing 
    relative radial distance, reaching 1 decade of overestimation 
    near $R_{200}$. The same trend but milder is present in the 
    less massive haloes.
    
    \item Although T11 model is able to fit individual profiles of
    RP in haloes with different masses and from different 
    epochs, a temporal or mass dependence of its free parameters
    can not be defined in order to set an universal analytic model 
    of RP. Moreover, by doing so, as in \cite{Tecce2011}, the analytic 
    profile obtained from the fitting processes tends to flatten 
    the predicted RP in radial distance, 
    missing the largest and homogeneous values of RP experienced
    by the satellites close to the centres of massive hosts. 
    
    \item We introduced a new analytic model for the RP profile
    which recovers the expected values measured from the simulation
    for a large range of main host halo masses, epochs and radial distances. 
    The profile, defined by the equation~\eqref{eq:pram_cvm}, 
    is characterised by a power which is a 
    function of the main host halo mass
    through \eqref{eq:pram_alpha}, i.e.
    \begin{equation}
            P_\textrm{ram} = P_0(z) 
             \left(\frac{r}{\xi(z) R_{200}} \right)^{-\frac{3}{2} \left[
                \alpha_\textrm{M}(z) \log (
        M_{200}\;h^{-1} [\textrm{M}_\odot] ) -5.5 \right] }, 
    \end{equation}
    and the temporal evolution of the three free parameters 
    were fit to the simulation RP measurements, giving 
    \begin{align}
     \log (P_0/(10^{-12}h^2\text{dyn\;cm}^{-2})) = 7.01 a^{-0.122} - 9.1, \\
     \xi = -3.4 a^{-0.42} + 10.2, \\
     \alpha_\text{M} = 3.3 \times 10^{-3} a^{1.33} + 0.512, 
    \end{align}
    expressed in terms of the scale factor $a$.
    
    \item By using a semi-analytic model of galaxy formation, we showed
    that the analytic fit proposed in this work to model the RP experienced 
    by satellite galaxies represent a significant improvement with respect
    to the T11 model, specially at high redshifts ($z \gtrsim 1$). 
    We note, however, that some systematic differences still persist when
    comparing with the model using RP values measured from the 
    underlying simulation. These differences are an expected limitation of 
    the RP modelling when using analytic profiles.
    
    \item In high-density environments like galaxy clusters, where 
    extreme high values of RP can be reached,  
    differences among the models of RP are larger with increasing 
    redshift. The comparison of the distribution of ratios
    between the accumulated stripped mass and the stellar mass of the 
    satellites shows an increasing reduction with increasing redshift 
    of the number of galaxies experiencing larger gas stripping 
    when the T11 model is applied.
    Moreover, satellites experiencing the largest amounts
    of total stripped mass 
    tend to have low stellar mass 
    ($M_\star \leq 10^{9.5} \text{M}_\odot$) 
    and their sSFR is dependent on the RP modelling applied, 
    particularly at high redshifts.
\end{itemize}

In summary, according to the analysis of galaxy properties of satellites 
being evolved using three variants to model RP affecting them, 
our new profile exhibits a large improvement in the overall treatment 
of this environmental process, specially at high redshifts ($z>1$).

\section*{Acknowledgements}

The authors acknowledge the anonymous reviewer 
for the valuable comments and suggestions
that contributed to improve the presentation of this research. 
CVM acknowledges support from ANID/FONDECYT through grant 3200918.
CVM and FAG acknowledges financial support from the 
Max Planck Society through a Partner Group grant.
FAG acknowledges support from ANID FONDECYT Regular 1181264. 
SAC acknowledges funding from {\it Consejo Nacional de Investigaciones
Cient\'{\i}ficas y T\'ecnicas} (CONICET, PIP-0387), 
{\it Agencia Nacional de Promoci\'on de la Investigaci\'on, 
el Desarrollo Tecnol\'ogico y la Innovaci\'on} 
(Agencia I+D+i, PICT-2013-0317, PICT-2018-03743), 
and {\it Universidad Nacional de La Plata} (G11-150), Argentina.
TH acknowledge CONICET, Argentina, for their supporting fellowships.

\section*{Data Availability}
Most of the data referred through this article are publicly
available in a dedicated repository in \textsc{Github} at
\url{https://github.com/cnvega/PramFits}.
It includes the codes and scripts developed for analysis
and figures, additional figures comparing both RP profiles with 
the measurements in each simulation snapshot, the
resulting parameter values and covariance matrices of the RP profile
fitting process, and the analysed galaxy catalogues of the three model 
variants. The raw data of the resimulations and 
the semi-analytic model of galaxy formation will be shared on 
reasonable request to the corresponding author.



\bibliographystyle{mnras}
\bibliography{references}






\bsp	
\label{lastpage}
\end{document}